\documentclass[aps,prb,twocolumn,showpacs,superscriptaddress]{revtex4}
\usepackage{bm,amsmath,amssymb,latexsym,mathrsfs,enumerate}
\usepackage[mathcal]{euscript}
\usepackage{hyperref}
\usepackage{graphicx}
\renewcommand{\vec}[1]{\bm{#1}}
\begin{document}

\date{January 27, 2005}

\title{Vortex motion in a finite--size easy--plane ferromagnet and application to a nanodot}

\author{Denis D.~Sheka}
    \email{Denis\_Sheka@univ.kiev.ua}
     \affiliation{National Taras Shevchenko University of Kiev, 03127 Kiev, Ukraine}

\author{Juan P.~Zagorodny}
    \affiliation{Physics Institute, University of Bayreuth, 95440 Bayreuth, Germany}

\author{Jean--Guy Caputo}
    \affiliation{ Laboratoire de Math\'ematiques, INSA de Rouen,
        B.P. 8, 76131 Mont-Saint-Aignan cedex France \\ and Laboratoire de
    Physique theorique et Modelisation, Universit\'e de Cergy-Pontoise
    and CNRS , 95031 Cergy-Pontoise cedex France }

\author{Yuri Gaididei}
    \affiliation{Institute for Theoretical Physics, 252143 Kiev, Ukraine}

\author{Franz G.~Mertens}
    \affiliation{Physics Institute, University of Bayreuth, 95440 Bayreuth, Germany}

\begin{abstract}

We study the motion of a non--planar vortex in a circular
easy--plane ferromagnet, which imitates a magnetic nanodot. Analysis
was done using numerical simulations and a new collective variable
theory which includes the coupling of Goldstone--like mode with the
vortex center. Without magnetic field the vortex follows a spiral
orbit which we calculate. When a rotating in--plane magnetic field
is included, the vortex tends to a stable limit cycle which exists
in a significant range of field amplitude $B$ and frequency $\omega$
for a given system size $L$. For a fixed $\omega$, the radius $R$ of
the orbital motion is proportional to $L$ while the orbital
frequency $\Omega$ varies as $1/L$ and is significantly smaller than
$\omega$. Since the limit cycle is caused by the interplay between
the magnetization and the vortex motion, the internal mode is
essential in the collective variable theory which then gives the
correct estimate and dependency for the orbit radius $R\sim B
L/\omega$. Using this simple theory we indicate how an ac magnetic
field can be used to control vortices observed in real magnetic
nanodots.

\end{abstract}


\maketitle

\section{Introduction}
\label{sec:intro}

Nonlinear topological excitations in 2D spin systems of soliton or
vortex type are known to play an essential role in 2D magnetism. For
example, solitons break the long--range order in 2D isotropic
magnets. Vortices play a similar role in 2D easy--plane magnets.
Magnetic vortices have been studied since the 1980s. They are
important for the dynamical and thermodynamical properties of
magnets, for a review see Ref.~\onlinecite{Mertens00}. The vortex
contribution to the response functions of 2D magnets has been
predicted theoretically\cite{Mertens89} and observed
experimentally\cite{Wiesler89}.

A second wind in the physics of magnetic vortices appeared less than
five years ago due to the direct observation of vortices in
permalloy (Py, $\mathrm{Ni}_{80}\mathrm{Fe}_{20}$)
\cite{Shinjo00,Cowburn98,Cowburn99,Pulwey01,Gubbiotti02,Park03} and
$\mathrm{Co}$ \cite{Fernandez00,Raabe00,Lebib01} magnetic nanodots.
Such nanodots are submicron disk--shaped particles, which have a
single vortex in the ground state due to the competition between
exchange and magnetic dipole--dipole interaction.\cite{Hubert98} A
vortex state is obtained in nanodots that are larger than a single
domain whose size is a few nanometers (e.g. for the Py--nanodot the
exchange length $l_{\text{ex}}=5.9\,nm$). The vortex state of
magnetic nanodots has drawn much attention because it could be used
for high-density magnetic storage and miniature sensors.
\cite{Cowburn02,Skomski03}. For this one needs to control
magnetization reversal, a process where vortices play a big role
\cite{Guslienko01}. The vortex signature has been probed by Lorentz
transmission electron microscopy\cite{Raabe00,Schneider02} and
magnetic force measurements\cite{Pokhil00,Fernandez00}. Great
progress has been made recently with the possibility to observe high
frequency dynamical properties of the vortex state magnetic dots by
Brillouin light scattering of spin waves
\cite{Demokritov01,Hillebrands02}, time--resolved Kerr microscopy
\cite{Park03}, phase sensitive Fourier transformation technique
\cite{Buess04}, and X--ray imaging technique \cite{Choe04}. These
have shown that the vortex performs a gyrotropic precession when it
is initially displaced from the center of the dot, e.g. by an
in--plane magnetic field pulse. \cite{Usov02,Guslienko02a,Park03}

In general the vortex mesoscopic dynamics is described by the Thiele
collective coordinate approach
\cite{Thiele73,Thiele74,Huber82,Nikiforov83}, which considers the
vortex as a rigid structure not having internal degrees of
freedom.\cite{Mertens00} However recent experimental and theoretical
studies
\cite{Ivanov98,Gaididei00,Raabe00,Pulwey01,Kovalev02,Kovalev03,Zagorodny03,Kovalev03a}
indicate phenomena which can not be explained using such a simple
picture. One striking example is the switching of the vortex
polarization
\cite{Gaididei00,Raabe00,Pulwey01,Kovalev02,Kovalev03,Zagorodny03},
where coupling occurs between the vortex motion and oscillations of
its core. Another one is the cycloidal oscillations of the vortex
around its mean path \cite{Ivanov98,Kovalev03a} where the dynamics
of the vortex center is strongly coupled to spin waves. In this way
the internal dynamics of the vortex plays a vital part. One of the
first attempts to take into account the internal structure of
vortices was presented in Ref.~\onlinecite{Caputo03} which showed
that a variation of the core radius slaved to the position explained
the motion of a vortex pair across an interface between two
materials of different anisotropy. Some progress has been achieved
in Ref.~\onlinecite{Zagorodny04} where we have confirmed that
internal degrees of freedom play a crucial role in the dynamics of
vortices driven by an external time-dependent magnetic field in a
classical spin system.

Here we present a complete study of this problem using direct
numerical simulations of the spin system and a collective variable
theory which includes an internal mode. We show that the periodic
forcing of the system by the time dependent magnetic field together
with the damping stabilizes the vortex in a finite domain. This
limit cycle exists because of the interplay between the
magnetization and the vortex position so that it is essential to
include an internal mode in the collective variable theory to
describe it. When this is done, the theory yields the domain of
stability in parameter space and the main dependencies on the field
amplitude $B$ and frequency $\omega$. It can be seen as a one of the
first generalizations to vortices of the collective variable
theories developed for 1D Klein-Gordon kinks by Rice
\cite{Rice83,Quintero00,Quintero00a} which include the width of the
kink together with its position.

In the next section \ref{sec:continuum} we formulate the continuum
model, discuss the role of different types of interactions and
briefly review the main results on the structure of the vortex
solution. The vortex motion without external field is examined in
section \ref{sec:nofield}.  It follows a spiral orbit as a result of
the competition between the gyroforce, the Coulomb force and the
damping force. In section \ref{sec:simulations} with the ac driving,
numerical simulations show that the vortex converges to a stable
limit cycle. We give its boundaries in parameter space and indicate
how the radius and frequency of the vortex orbital motion depends on
the field and geometry parameters. Section \ref{sec:field} presents
and discusses in detail the \emph{new collective variable theory} of
the observed vortex dynamics which takes into account the coupling
between an internal shape mode and the translational motion of the
vortex position. In section \ref{sec:discussion} we link this with
the individual spin motion observed in the simulations and indicate
how these effects can be observed and used in real nano magnets.

The model we consider is a ferromagnetic system with spatially
homogeneous uniaxial anisotropy, described by the classical
Heisenberg Hamiltonian
\begin{equation} \label{eq:H-discrete}
\begin{split}
\mathscr{H}_0 = &-\frac{J}{2}\!\!
\sum_{\left(\vec{n},\vec{n}'\right)}\! \left(\vec{S}_{\vec{n}}\cdot
\vec{S}_{\vec{n}'} - \delta S^z_{\vec{n}} S^z_{\vec{n}'} \right)
 + \frac{K}{2}\!\! \sum_{\vec{n}} \left(S^z_{\vec{n}} \right)^2\!\! .
\end{split}
\end{equation}
Here  $\vec{S}_{\vec{n}}\equiv\left(S^x_{\vec{n}}, S^y_{\vec{n}},
S^z_{\vec{n}}\right)$ is a classical spin vector with fixed length
$S$ on the site $\vec{n}$ of a two-dimensional square lattice, and
the exchange integral $J > 0$ for a ferromagnet. The first summation
runs over nearest--neighbor pairs $(\vec{n},\vec{n}')$. We assume a
small anisotropy leading to an easy-plane ground state. This
anisotropy can be either of the exchange type, with $0<\delta\ll 1$,
or of the on--site type, with $0 \leq K \ll J$.

Extending ideas of Ref.~\onlinecite{Zagorodny04} we study the
movement of a vortex in this system under the action of a magnetic
field $\vec{B}(t)=\left(B\cos\omega t, B\sin\omega t, 0\right)$,
which is spatially homogeneous and is rotating in the plane of the
lattice. This field adds an interaction of the form
\begin{equation} \label{eq:V(t)-discrete}
\mathscr{V}(t) = -\gamma B \sum_{\vec{n}}\left( S_{\vec{n}}^x \cos
\omega t + S_{\vec{n}}^y \sin \omega t \right),
\end{equation}
where $\gamma=2\mu_B/\hslash$ is the gyromagnetic ratio.

The spin dynamics is described by the Landau--Lifshitz equations
with Gilbert damping
\begin{equation} \label{eq:LL-discrete}
\frac{d \vec{S}_{\vec{n}} }{dt} =  - \left[\vec{S}_{\vec{n}}\times
\frac{\partial \mathscr{H} }{\partial \vec{S}_{\vec{n}}}\right] -
\frac{\varepsilon}{S} \left[\vec{S}_{\vec{n}} \times \frac{d
\vec{S}_{\vec{n}} }{dt}\right],
\end{equation}
where $\mathscr{H}=\mathscr{H}_0+\mathscr{V}(t)$ is the total
Hamiltonian. Eqs.~\eqref{eq:LL-discrete} preserve the length of the
spins $|\vec{S}_{\vec{n}}| \equiv S $,  which has units of action.
Another form of Eqs.~\eqref{eq:LL-discrete} more suitable for spin
dynamics simulations is given in Appendix
\ref{sec:appendix-Discrete}.

\section{Continuum Limit}
\label{sec:continuum}

In the case of weak anisotropies $\delta \ll  1$, $K \ll J $, the
characteristic size of excitations $l_0 = a \sqrt{{J}/({ 4 J
\delta+K })}$  is larger than the lattice constant $a$, so that in
the lowest approximation on the small parameter $a/l_0$ and weak
gradients of magnetization we can use the continuum approximation
for the Hamiltonian \eqref{eq:H-discrete}
\begin{equation} \label{eq:H-continuum}
H_0 \equiv  \mathscr{H}_0 - E_0  =  \frac{J S^2}{2} \int d^2 x
\left[ \left(\nabla \vec{s} \right)^2 +  \frac{m^2}{l_0^2} \right],
\end{equation}
where $E_0$ is a constant. The spin length has been rescaled so that
\begin{equation} \label{eq:s}
\vec{s} = \vec{S}/S =
\left(\sqrt{1-m^2}\cos\phi;\sqrt{1-m^2}\sin\phi;m\right)
\end{equation}
is a unit vector. The length $l_0$ coincides with the radius of the
vortex core obtained in Ref.~\onlinecite{Nikiforov83} for on-site
anisotropy type alone ($\delta =0$). For the case of exchange
anisotropy alone ($K=0$), it is also customary to use the length
$r_{\text{v}} = a \sqrt{(1-\delta)/4 \delta},$ which is obtained
from an asymptotic analysis and is to be identified later with the
radius of the ``core'' of a vortex.\cite{Gouvea89a,Mertens97}
However, for the range of $\delta$ we are interested in, \emph{i.e.}
for $\delta \lesssim0.1$,  the difference between $r_\text{v}$ and
$l_0$ is negligible.

The interaction with a homogeneous time-dependent magnetic field is
expressed as
\begin{equation}\label{eq:V-continuum}
\begin{split}
V(t) &= -\frac{J S^2}{l_0^2}\, \int\, d^2x \left(\vec{b}(t) \cdot \vec{s}(\vec{r},t)\right) \\
&= -J S^2b \int\,\,
d^2\xi\,\,\sqrt{1-m^2}\cos\left(\phi-\nu\tau\right).
\end{split}
\end{equation}

In order to simplify notations we use here and below the
dimensionless coordinate $\vec{\xi}\equiv \vec{r}/l_0$, the
dimensionless time $\tau\equiv\omega_0 t$, the dimensionless driving
frequency $\nu=\omega/\omega_0$ and the dimensionless magnetic field
$\vec{b}=\gamma\vec{B}/\omega_0$,\cite{Ivanov95b,Ivanov02} where
\begin{equation} \label{eq:omega0}
\omega_0 =  S\left(4J\delta + K\right).
\end{equation}

In all real magnets there is, in addition to short--ranged
interactions, a long--ranged dipole--dipole interaction. In the
continuum limit this interaction can be taken into account as energy
of an effective demagnetization field, $\vec{H}^{(m)}$
\begin{equation*} \label{eq:E-demagnetization}
\mathscr{E}^{(m)}  = -  \int\! d^2 x\ \vec{M}\cdot\vec{H}^{(m)},
\end{equation*}
where $\vec{M}$ is the magnetization. Generally, this field is a
complicated functional of $\vec{M}$. However, in the case of a thin
magnetic film (or particle) the volume contribution to the
demagnetization field is negligible, and only surface fields are
important. The face surfaces produce a local field
$\vec{H}^{(m)}=-4\pi M_0\vec{e}_z$ for the sample with the
saturation magnetization $M_0$. Then the dipole--dipole interaction
can be taken into account by a simple redefinition of the anisotropy
constants, $K\to K^{\text{eff}}=K+4\pi M_0^2a^2/S^2$, leading to a
new magnetic length \cite{Ivanov01}
\begin{equation} \label{eq:l0-eff}
l_0\to l_0^{\text{eff}} = a\sqrt{\frac{J}{4J\delta+K+4\pi
M_0^2a^2/S^2}}.
\end{equation}
This is the case of so--called configurational or shape anisotropy.
\cite{Schabes88,Cowburn98,Cowburn02} The lateral surface affects
only the boundary conditions, see
Refs.~\onlinecite{Guslienko02b,Ivanov02a} for details. For example,
for a very thin magnetic particle, which corresponds to our 2D
system, free boundary conditions are valid, and we will use them in
the paper.

Thus, the total energy functional, normalized by $JS^2$, reads
\begin{equation} \label{eq:E-total}
\mathscr{E}[\vec{s}]  = \int d^2\xi
\left[\frac{\left(\nabla\vec{s}\right)^2}{2}+\frac{m^2}{2}
-\left(\vec{b}\cdot\vec{s}\right)\right],
\end{equation}
where we have rescaled the magnetic length in accordance with
\eqref{eq:l0-eff}.

The continuum version of the Landau--Lifshitz
Eqs.~\eqref{eq:LL-discrete} becomes
\begin{subequations} \label{eq:LL-continuum}
\begin{align} \label{eq:LL-continuum(1)}
\frac{\partial \phi}{\partial \tau} &=
\frac{\delta\mathscr{E}}{\delta m}
+ \frac{\varepsilon}{(1-m^2)}\frac{\partial m}{\partial \tau}, \\
\label{eq:LL-continuum(2)} \frac{\partial m}{\partial \tau}  &= -
\frac{\delta\mathscr{E}}{\delta\phi} -
\varepsilon(1-m^2)\frac{\partial \phi}{\partial \tau}.
\end{align}
\end{subequations}
These equations can be derived from the Lagrangian
\begin{equation} \label{eq:Lagrangian}
\mathscr{L} = -\int d^2\xi  (1-m)\frac{\partial\phi}{\partial \tau}
- \mathscr{E}[\vec{s}]
\end{equation}
and the dissipation function
\begin{align}\label{eq:F-diss}
\mathscr{F} &=  \frac{\varepsilon}{2} \int d^2\xi \left(\frac{\partial\vec{s}}{\partial \tau}\right)^2\\
&= \frac{\varepsilon}{2}\int d^2\xi \left[\frac{1}{1-m^2}
\left(\frac{\partial m}{\partial\tau}\right)^2 + (1-m^2)
\left(\frac{\partial \phi}{\partial \tau}\right)^2\right]. \nonumber
\end{align}
Then Eqs.~\eqref{eq:LL-continuum} result explicitly in
\begin{subequations} \label{eq:LL-m-phi}
\begin{align} \label{eq:LL-m-phi(1)}
\frac{\partial\phi}{\partial \tau} &= -\frac{m\, \left(\vec{\nabla}
m\right)^2}{(1-m^2)^2} + m\left[ 1 - \left(\vec{\nabla}
\phi\right)^2\right] -\frac{ \Delta m}{1-m^2}
\nonumber\\
&+ \frac{b\, m\, \cos(\phi - \nu\tau)}{\sqrt{1 -m^2}}
+ \frac{\varepsilon}{(1-m^2)}\, \frac{\partial m}{\partial \tau}, \\
\label{eq:LL-m-phi(2)} \frac{\partial m}{\partial \tau} &=
\vec{\nabla}[ (1 -m^2)\vec{\nabla}\phi ] - b \sqrt{1 -m^2} \sin(\phi
- \nu \tau)
\nonumber\\
&- \varepsilon (1-m^2) \frac{\partial \phi}{\partial \tau}.
\end{align}
\end{subequations}

Without magnetic field the ground state of the system is a uniform
planar state $m=0$ and $\phi=\text{const}$. The field changes
essentially the picture: spins start to precess homogeneously in the
XY--plane, $\phi=\varphi+\nu\tau$. Such a precession causes the
appearance of a z--component of magnetization, $m=\text{const}$.
From Eqs.~\eqref{eq:LL-m-phi}, we find that the equilibrium values
of $m$ and $\phi$ satisfy the following equations,
\begin{subequations} \label{eq:ground}
\begin{align}
\label{eq:ground(m0)}
& \left(1-\frac{\nu}{m}\right)^2 + \varepsilon^2\nu^2 = \frac{b^2}{1-m^2},\\
\label{eq:ground(phi0)} & -b\sin \phi -
\varepsilon\nu\sqrt{1-m^2}=0,
\end{align}
\end{subequations}
so that this state can only exist if $b \geq \varepsilon\nu$
(otherwise only the ground state with $m=0$ and $\phi=\text{const}$
exists). Assuming $m\ll 1$, we obtain
\begin{equation} \label{eq:m0}
m\approx\frac{\nu}{1 - \sqrt{b^2-\varepsilon^2\nu^2}}, \quad \phi =
\nu\tau +\pi +  \arcsin \frac{\varepsilon\nu\sqrt{1-m^2}}{b}.
\end{equation}
Note how the magnetization $m$ is proportional to the field
frequency $\nu$ so that its sign is important. Below we discuss the
role of this homogeneous solution in the vortex dynamics.

The continuum analogue of the power--dissipation relation
\eqref{eq:dH/dt} for the total energy functional
$\mathscr{E}[\vec{s}]$ is calculated from Eqs.
\eqref{eq:Lagrangian}, \eqref{eq:F-diss} and gives
\begin{equation} \label{eq:dE/dt}
 \frac{d \mathscr{E}}{d\tau} = - 2 \mathscr{F} - \mathscr{W},\qquad
\mathscr{W} =  \int d^2 \xi
\left(\vec{s}\cdot\frac{d\vec{b}}{d\tau}\right).
\end{equation}
Formally, Eqs.~\eqref{eq:ground} have two solutions. One can check
that only for the solution  \eqref{eq:m0} the dissipation balances
the work done by the field, so that the energy $\mathscr{E}$ tends
to be stabilized.

\subsection*{Static Vortices}
\label{sec:static}

The simplest nonlinear excitation of the system is the well--known
non--planar magnetic vortex. We recall briefly  the structure of a
single static vortex at zero field. In this case the pair of
functions $(m,\phi)$ satisfies the Eqs.~\eqref{eq:LL-continuum} with
the time derivatives set to zero and $b=0$. If we look for planar
solutions ($m=0$) for the $\phi$ field,
Eq.~\eqref{eq:LL-continuum(2)} becomes the Laplace equation. For the
vortex solution located at $Z=X+iY=R\exp(i\Phi)$ the $\phi$--field
has the form:
\begin{equation} \label{eq:phi-static}
\phi(z)=\varphi_0+q\arg\left(z-Z\right),
\end{equation}
where $z=x+iy$ is a point of the XY--plane, $q\in\mathbb{Z}$ is the
$\pi_1$ topological charge of the vortex (vorticity). We will call
the solution with $q=1$ a vortex and the solution with $q=-1$ an
antivortex. The expression \eqref{eq:phi-static} does not satisfy
the boundary conditions for a finite system. For our circular system
of radius $L$ (in units of $l_0$) and free boundary conditions the
solution  is \cite{Kovalev03a}
\begin{equation} \label{eq:phi-staticV+AV}
\phi=\arg\left(z-Z\right) - \arg\left(z-Z_I\right) + \arg Z,
\end{equation}
where the ``image'' vortex is added at $Z_I=ZL^2/R^2$ to satisfy the
Neuman boundary conditions. The last term in
\eqref{eq:phi-staticV+AV} is inserted to have the correct limit for
$L \rightarrow \infty$.

The $m$--field has radial symmetry,
$m\equiv\cos\theta\left({\rho}\equiv|z-Z|\right)$. From
\eqref{eq:LL-m-phi(1)} and \eqref{eq:phi-static} one can derive that
$\theta(\bullet)$ satisfies the following differential problem:
\begin{subequations} \label{eq:m-static}
\begin{align}
\label{eq:m-static(1)}
&\frac{d^2\theta}{d{\rho}^2}+\frac1\rho\frac{d\theta}{d{\rho}}+\sin\theta\cos\theta
\left(1-\frac{1}{{\rho}^2}\right)=0,\\
\label{eq:m-static(2)} &\cos\theta(0)=p, \qquad
\cos\theta(\infty)=0,
\end{align}
\end{subequations}
where $p=\pm 1$ is the so--called polarity of the vortex. The
solution of this differential problem is a bell--shaped structure
with a width in the order of $l_0$.

\section{Vortex motion at zero field}
\label{sec:nofield}

A standard description for the steady movement of magnetic
excitations was given first by Thiele. \cite{Thiele73,Thiele74}
Huber \cite{Huber82}, Nikiforov and Sonin \cite{Nikiforov83} have
first applied this approach to the dynamics of magnetic vortices,
using a traveling wave Ansatz $\vec{s}(z,\tau) =
\vec{s}\left(z-Z(\tau)\right)$. In terms of the fields $m$ and
$\phi$ such an Ansatz is
\begin{subequations} \label{eq:Thiele-ansatz}
\begin{align}
\label{eq:Thiele-ansatz(1)} %
m(z,\tau)   &= \cos\theta\left(|z-Z(\tau)|\right),\\
\label{eq:Thiele-ansatz(2)} %
\phi(z,\tau) &= \arg\left(z-Z(\tau)\right) -
\arg\left(z-Z_I(\tau)\right) + \arg Z(\tau),
\end{align}
\end{subequations}
where the function $\theta(\bullet)$ describes the out--of--plane
structure of the static vortex, and is the solution of
Eqs.~\eqref{eq:m-static}.

To derive an effective equation of the vortex motion for the
collective variable $\vec{R}(\tau)=\left(X(\tau),Y(\tau)\right)$, we
project the Landau--Lifshitz Eqs.~\eqref{eq:LL-continuum} over the
lattice using Ansatz \eqref{eq:Thiele-ansatz}. We obtain a Thiele
equation in the form of a force balance,\cite{Mertens00}
\begin{equation} \label{eq:Thiele}
G\left[\vec{e}_z\times\dot{\vec{R}} \right] - 2 \pi \eta
\dot{\vec{R}} + \vec{F}=0,
\end{equation}
where the dot indicates derivative with respect to the rescaled time
$\tau$. The first term, the gyroscopic force, acts on the moving
vortex and determines the main properties of the vortex dynamics.
The value of the gyroconstant is well--known $G=2\pi pq$,
\cite{Huber82,Nikiforov83} in our case for the vortex with positive
polarity and unit vorticity $G=2\pi$. The second term describes the
damping force with a coefficient \cite{Huber82,Kamppeter99}
\begin{equation} \label{eq:eta}
\eta =  {1 \over 2} \varepsilon (\ln {L} + C_1),
\end{equation}
where $C_1\approx 2.31$ is a constant coming from the $m$ field and
is calculated in the appendix, see formula \eqref{eq:C1}. The $\ln
{L}$ dependence in $\eta$ was obtained in Ref.~\onlinecite{Huber82}.

The last term in Eq.~\eqref{eq:Thiele} is an external force, acting
on the vortex, $\vec{F} = -\vec{\nabla}_{\!\!\vec{R}} \mathscr{E}$,
where $\mathscr{E}$ is the total energy functional
\eqref{eq:E-total}. Without magnetic field ($\vec{b}=0$) such a
force appears as a result of boundary conditions, it describes the
2D Coulomb interaction between the vortex and its image
\begin{equation} \label{eq:E-int}
\mathscr{E}^{\text{int}} = \mathscr{E}_0 + \pi \ln\frac{L^2-R^2}{L},
\end{equation}
where $\mathscr{E}_0 \approx \pi$ is the energy of the vortex
core.\cite{Kovalev03a}

In order to generalize the effective equations of the vortex motion
for the case of the magnetic field we derive now the same effective
equations by the effective Lagrangian technique as it was proposed
in Refs.~\onlinecite{Zagorodny03,Caputo03,Zagorodny04}. Inserting
Ansatz \eqref{eq:Thiele-ansatz} into the ``microscopic'' Lagrangian
\eqref{eq:Lagrangian} and the dissipative function
\eqref{eq:F-diss}, and calculating the integrals, we derive an
effective Lagrangian (see Appendix \ref{sec:appendix-nofield} for
the details)
\begin{equation} \label{eq:L-eff}
\mathscr{L} = -\pi R^2\dot{\Phi} - \mathscr{E}^{\text{int}}.
\end{equation}
In the same way we derive the effective dissipative function
\begin{equation} \label{eq:F-eff}
\mathscr{F} = \pi \eta\dot{\vec{R}}^2= \pi \eta\left(\dot{R}^2 +
R^2\dot{\Phi}^2\right).
\end{equation}
The equations of motion are then obtained from the Euler--Lagrange
equations
\begin{equation} \label{eq:Euler-Lagrange}
\frac{\partial \mathscr{L}}{\partial X_i}  - \frac{d \phantom{t}}{d
\tau}  \left(\frac{\partial\mathscr{L}}{\partial \dot{X}_i}\right) =
\frac{\partial \mathscr{F}}{\partial \dot{X}_i}
\end{equation}
for the $X_i = \left\{R,\Phi\right\}$,
\begin{subequations} \label{eq:CV-equation-no-field}
\begin{align}
\label{eq:CV-equation-no-field(1)}
&\dot{\Phi} + \eta\frac{\dot{R}}{R} = \frac{1}{L^2-R^2},\\
\label{eq:CV-equation-no-field(2)} &\frac{\dot{R}}{R} = \eta
\dot{\Phi}.
\end{align}
\end{subequations}
This set of equations is equivalent to the Thiele
Eq.~\eqref{eq:Thiele}, when going to polar coordinates.

For zero damping ($\varepsilon=0$) two radial forces act on the
vortex (gyroforce and Coulomb force) and compensate each other,
providing pure circular motion of the vortex. In that case the
radius $R$ of the orbit is arbitrary. Using
Eqs.~\eqref{eq:CV-equation-no-field} it is easy to calculate the
frequency of this circular motion for a given $R$, see
Ref.~\onlinecite{Mertens00}:
\begin{equation} \label{eq:Omega_0}
\Omega (R) = \frac{1}{L^2-R^2}.
\end{equation}

When the damping is present, there appears an additional damping
force which can not be compensated by other forces. Thus the
trajectory of the vortex becomes open--ended, following the
logarithmic spiral from \eqref{eq:CV-equation-no-field(2)}:
\begin{equation} \label{eq:spiral-trajectory}
\Phi-\Phi_0 = \frac{1}{\eta}\ln\frac{R}{R_0},
\end{equation}
where $R_0$ and $\Phi_0$ are constants.

\section{Numerical simulations of the vortex dynamics}
\label{sec:simulations}

\begin{figure}
\includegraphics[width=\linewidth]{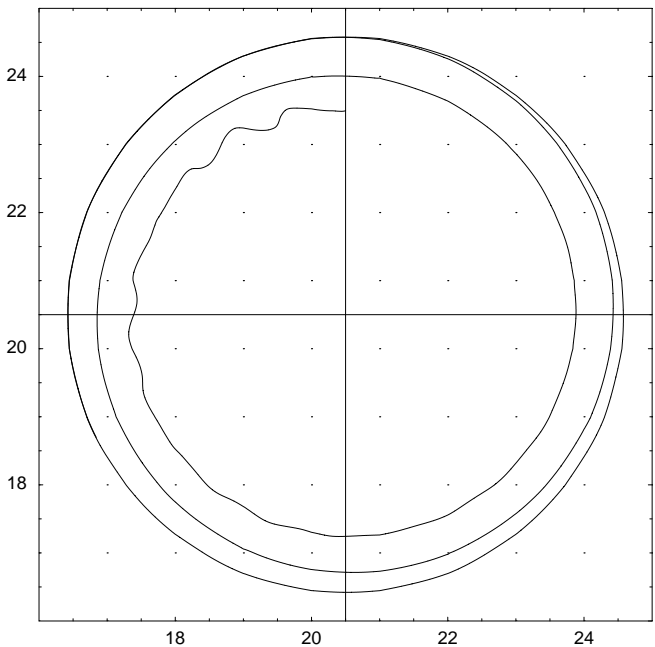}
\caption{~Trajectory of a vortex at zero field for a time interval
$0<t<10^4$. The damping $\varepsilon=0.01$ was switched off at
$t=1600$. The vortex with $q=p=1$ was launched from \mbox{$Z = 20.5a
+ i23.5a$} on a lattice of radius $L=20a\approx 11.3$. ``Clean''
circular trajectories, where the vortex is free of spin waves,  are
obtained with this method. In the whole study the anisotropy is set
to $\delta=0.08$. The damping is $\varepsilon=0.01$. }
\label{fig:trajecs}
\end{figure}

To investigate the vortex dynamics, we integrate numerically the
discrete Landau--Lifshitz equations \eqref{eq:LL-discrete(num)} over
square lattices of size $(2L)^2$ using a 4th--order Runge--Kutta
scheme with time step 0.01. Each lattice is bounded by a circle of
radius $L$ on which the spins are free corresponding to a Neuman
boundary condition in the continuum limit. In all cases the vortex
is started near the center of the domain and the field and damping
are turned on adiabatically over a time interval of about 100. We
have only considered vortices of fixed polarity $p=1$. More details
on the numerical procedure and in particular the vortex tracking
algorithm can be found in Ref.~\onlinecite{Zagorodny03}.

We have fixed the exchange constant $J=1$ as well as the spin length
$S=1$. All cases presented here are for the anisotropy
$\delta=0.08$, corresponding to $l_0 \approx 1.77 a$ so that we are
close to the continuum limit. The lattice radii we consider here are
$20a <L <100a$.

To validate the simple theory presented in the previous section we
considered the case with no magnetic field. In the absence of
damping the vortex should follow a circular orbit and its frequency
of rotation should be given by \eqref{eq:Omega_0}. Starting with a
vortex initial condition for $m$ and $\phi$ given by
\eqref{eq:Thiele-ansatz}, it is possible to ``prepare'' circular
trajectories of arbitrary radius by applying damping. This kills all
spin waves coming from the imperfect initial condition and drives
the vortex to the selected radius following the spiral
\eqref{eq:spiral-trajectory}. Once the chosen radius is reached,
damping is turned off \emph{adiabatically} over a time greater than
100 ($1 /\varepsilon$) and the vortex will keep its circular orbit
indefinitely. Such a scenario is shown in Fig.~\ref{fig:trajecs}.

\begin{figure}
\includegraphics[width=\columnwidth]{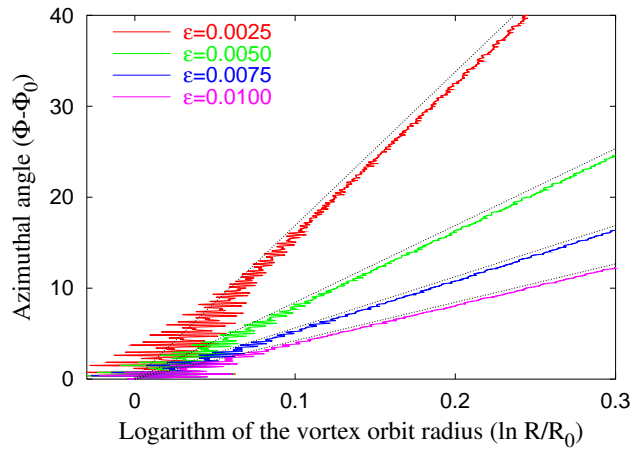}
\caption{ (Color online) Azimuthal angle $\Phi$ of the vortex
position as a function of the logarithm of the radial position $R$
for four different values of the damping $\varepsilon$. The lattice
radius is $L=20a \approx 11.3$. } \label{fig:PhiLogR}
\end{figure}

We now analyze the spiral trajectories obtained when damping is
present. In Fig.~\ref{fig:PhiLogR} we plot the measured angle of
rotation $\Phi$ (in radians) as a function the logarithm of the
measured radius $R$ for four values of damping. The vortex is
started every time from the same place ($\Phi_0=\pi/2$, $R_0=3a $ )
in the lattice. The behavior given by the spin simulation shown by
full lines agrees well with the relation
\eqref{eq:spiral-trajectory} given by dashed lines. Note that the
constant $C_1$ is important to obtain a quantitative agreement
because it is of the same order as the term $\ln L$.

\begin{figure}
\includegraphics[width=\columnwidth]{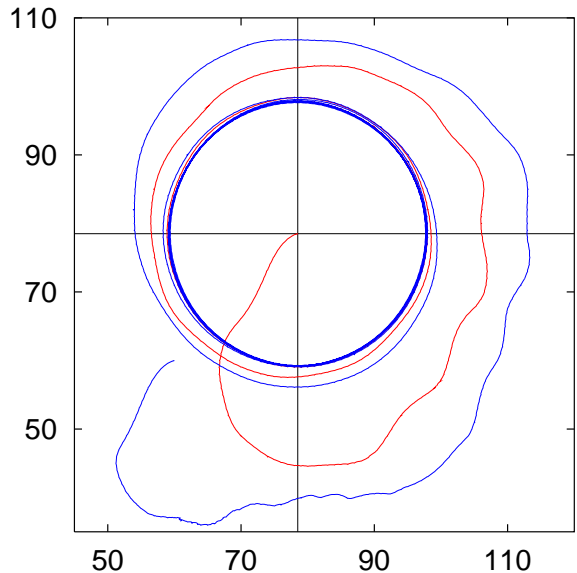}
\caption{ (Color online) Two trajectories of a vortex from simulations of the many-spin model
\eqref{eq:H-discrete}--\eqref{eq:LL-discrete}, on a lattice of
radius $L=78a \approx 44$,  with a rotating field %
($\nu=0.125$, $b=0.002$). For this field, all
trajectories converge to the same circle independently of the
vortex's initial position,  provided it is not too close to the
system border.
\label{fig:Rinit0} %
}
\end{figure}

To study the vortex dynamics in the presence of the rotating field,
we extend the simulations described in
Ref.~\onlinecite{Zagorodny03}. There we investigated the dynamics of
the out--of--plane structure of the vortex, focusing on the
phenomenon of switching, which occurs when $\nu p < 0 $. Here we
consider vortices with positive polarity $p=1$ and $\nu>0$ so that
no switching occurs.

For simplicity we fixed the damping $\varepsilon=0.01$ in
\eqref{eq:LL-discrete} and varied the parameters $(b,\nu,L)$. We
checked that the effects reported here occur for a range of
anisotropies and damping around these values. Given a combination of
the parameters ($\nu$,$b$) of the field, the radius $L$ of the
system and the damping $\varepsilon$, we have observed that either
the vortex escapes from the system through the border or it stays
inside for all times. In the latter case, it can approach a limit
cycle for a broad range of the field parameters.
Fig.~\ref{fig:Rinit0} shows two vortex trajectories starting from
different positions and converging to the same circle. When the
limit cycle exists, its basin of attraction is very large as can be
seen by starting the vortex at different positions and seeing it
converge to the same circle. In other words, the system keeps no
memory of the initial position of the vortex.

\begin{figure}
\centering
\includegraphics[width=\linewidth]{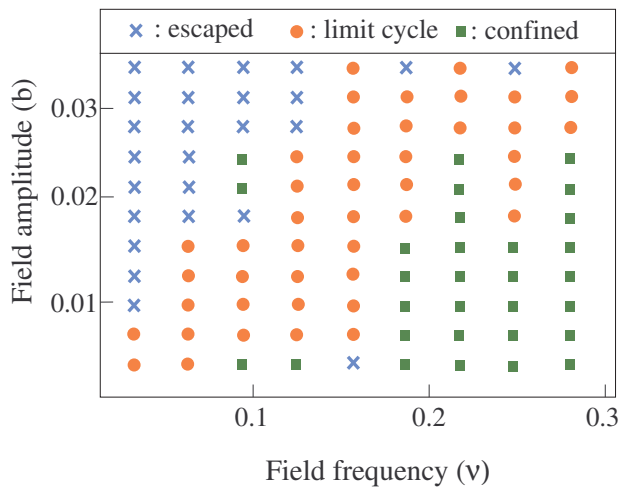}
\caption{ (Color online) Diagram of types of trajectories in the
$(\omega,B)$ parameter plane corresponding in the $(\nu,b)$ plane to
the range $0<\nu<0.3 $ and $0<b<0.033$. The radius of the system is
$L=36a\approx 20$. The term ``confined'' means that no limit cycle
was reached, though the vortex stayed inside the system, during the
time of observation $\tau \lesssim 6400$. %
\label{h-w-diagram}}
\end{figure}

To exist, the limit cycle  needs both magnetic field and damping:
once it is attained,  switching off or changing either of them
destroys immediately the circular trajectory. For fixed $\nu$ and
$L$, when the intensity $b$ is not large enough, damping dominates
and the vortex escapes from the system following a spiral, as
explained in the previous Section. If $b$ is too large, the vortex
will also escape due to an effective drift force caused by the
field,  which changes its direction slowly enough, relative to the
movement of the vortex.  This is the case when the frequency is very
small, such that the field is practically static. If both the
intensity and the frequency are too large, the field will destroy
the excitation creating many spin waves and also new vortices can be
generated from the boundary. Many seemingly chaotic trajectories can
be observed for high values of field parameters. To determine the
limit cycle, the value of the damping is not as critical as the
field parameters. For example, increasing the damping up to five
times its value ($\varepsilon=0.002$ to $0.01$) did not change
significantly the limit cycle shown in Fig.~\ref{fig:Rinit0} but
only accelerated the reaching of it. At this point note that there
is \emph{no resonant absorption} of the energy in the ac field
unlike the predictions of Ref.~\onlinecite{Usov02}. The field just
drives the vortex with the frequency $\Omega$, which is always lower
than the frequency $\omega$ of the ac field.

All these extreme cases constrain the size and shape of the regimes
where circular limit--trajectories appear in the space of field
parameters $(\nu,b)$. In Fig.~\ref{h-w-diagram} we show for a system
radius $L=36a$ this parameter plane and point out where the vortex
escapes or gives rise to a limit cycle or confined orbit. Similarly
to what we found in the study of switching, \cite{Zagorodny03}  we
also find ``windows'', \emph{i.e.} events which are not expected in
a particular region (for instance, the point $(\nu=0.1,b=0.02)$ in
the diagram). The zoom--in of any region of the diagram containing
windows shows again a similar behavior. We can also observe that the
vortex is sensitive to small variations of the field parameters, and
that its behavior is not monotonous (follow for example the line
$b=0.025$ for increasing frequencies).

When $L$ is varied, there can appear ``windows'' where there is no
limit cycle. For example for $L=36a$, $\nu=0.1$, $b=0.02$ the vortex
escapes from the system, while for $L=24a,30a$, on one side and
$L=42a,48a,54a,\cdots$ , on the other side, the vortex reaches a
limit cycle.

\begin{figure}
\centering
\includegraphics[width=\linewidth]{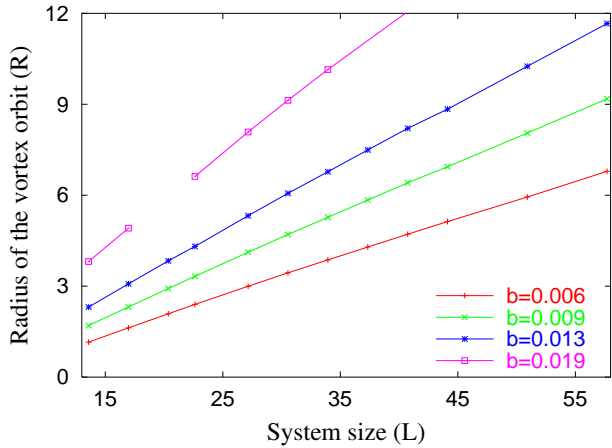}
\caption{ (Color online)  Radius of the vortex orbit $R$  \emph{
vs.} the system radius $L$, in the circular limit cycle, for a fixed
field frequency $\nu=0.094$ and several amplitudes $b$. The lines
are there to guide the eye. Here and in the next figures $R$ and $L$
are given in units of $l_0$. \label{fig:Rv-L}}
\end{figure}

\begin{figure}
\centering
\includegraphics[width=\linewidth]{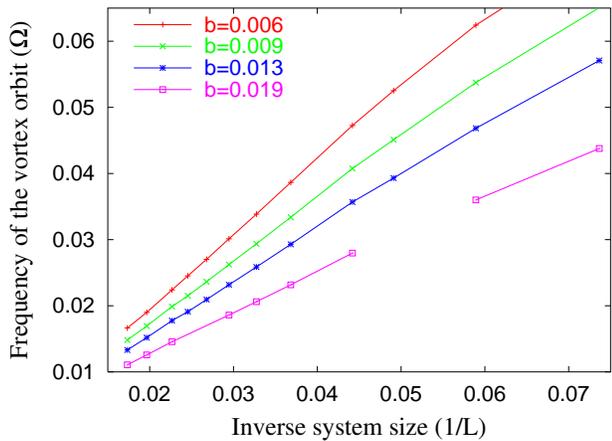}
\caption{ (Color online) Frequency of the vortex orbit $\Omega$
\emph{vs.} the inverse system radius $1/L$, in the circular limit
cycle, for a fixed field frequency $\nu=0.094$ and several
amplitudes $b$. \label{fig:Wv-L}}
\end{figure}
\begin{figure}
\centering
\includegraphics[width=\linewidth]{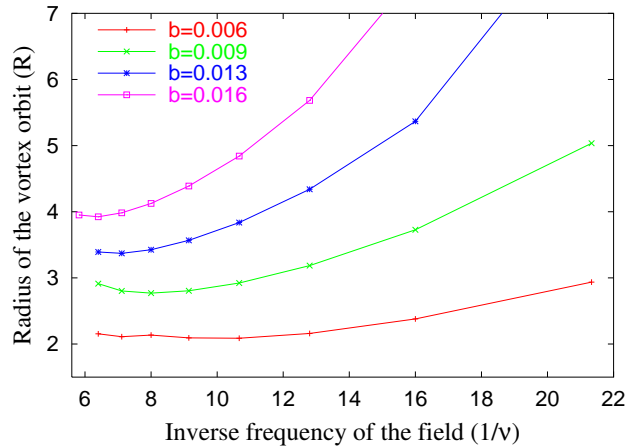}
\caption{ (Color online)  Radius of the vortex orbit $R$  \emph{
vs.} the inverse $1/\nu$ of the frequency $\nu$ of the rotating
magnetic field for four different amplitudes of the field. The
radius of the system is $L=36a\approx 20$. \label{fig:Rv-nu}}
\end{figure}

\begin{figure}
\centering
\includegraphics[width=\linewidth]{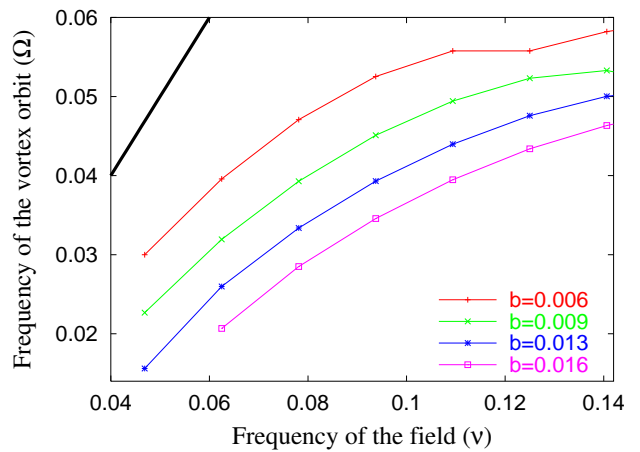}
\caption{ (Color online) Frequency of the vortex orbit $\Omega$
\emph{vs.} the frequency $\nu$ of the rotating magnetic field for
four different amplitudes of the field. The diagonal is plotted in
the upper left corner. The radius of the system is $L=36a\approx
20$. \label{fig:Wv-nu}}
\end{figure}

In the rest of this work we will concentrate on the circular limit
cycle.
Figs.~\ref{fig:Rv-L} and \ref{fig:Wv-L} show the dependence of the
vortex radial position $R$ and azimuthal frequency $\Omega$ as a
function of the system radius $L$ for a fixed field frequency
$\nu=0.094$ and four amplitudes $b$. The linear dependence of $R$ on
the system size $L$ is very clear from Fig.~\ref{fig:Rv-L} for the
whole range $11< L< 56$. Fig. \ref{fig:Wv-L} shows the frequency
$\Omega$ of the vortex orbit as a function of $1/L$. The dependence
is linear for $L>30$ but not for smaller $L$ indicating a possible
size effect. The points missing in the two figures for $L=20$ and
$b=0.0187$ correspond to a vortex escaping from the system.

For a fixed system size $L$ the features of the limit cycle depend
on the values of the field frequency $\nu$ and amplitude $b$. In
Fig.~\ref{fig:Rv-nu} we plot the radius $R$ of the limit cycle as a
function of the inverse $1/\nu$ of the frequency of the applied
field. For large frequencies one can see that the radius tends to a
constant which is proportional to the amplitude $b$. For low
frequencies the radius increases sharply. In this case damping plays
a larger role than mentioned above.

In Fig.~\ref{fig:Wv-nu} we plot the frequency $\Omega$ of the
orbital motion of the vortex as a function of the field frequency
$\nu$ for four values of the field amplitude $b$. The diagonal is
shown on the upper left corner of the picture and indicates that
$\Omega \ll \nu$.

\begin{figure}
\centering{a) $\nu=0.25,\; b =0.0025$}
\includegraphics[width=0.42\textwidth]{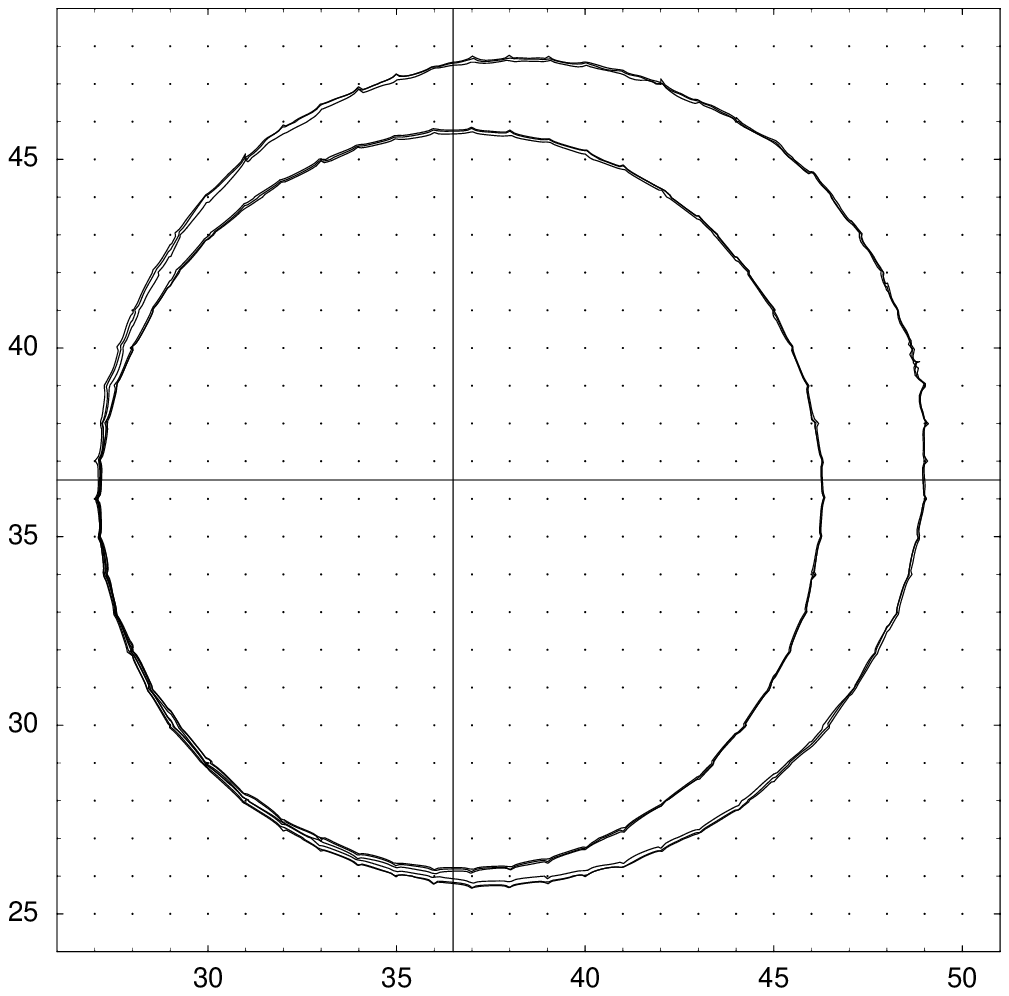}\\
\centering{b) $\nu=0.28,\; b=0.016$}
\includegraphics[width=0.42\textwidth]{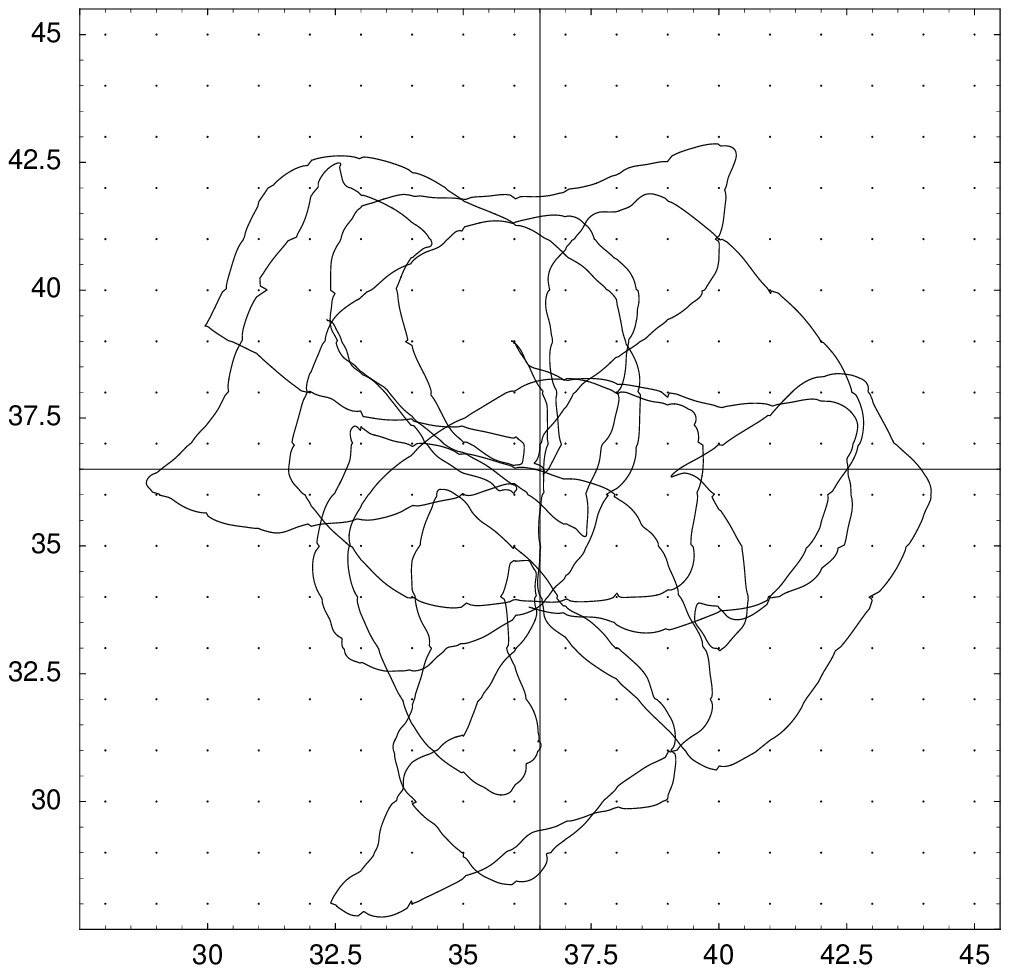}
\caption{Two different kinds of confined vortex trajectories that
are not circular, occurring for large field amplitudes and
frequencies. In the top panel the radial position $R(t)$ of the
vortex is periodic while it is chaotic in the bottom panel
\label{fig:conf-trajs}}
\end{figure}

Although most trajectories which converge to limit cycles end up in
a circular orbit around the center of the system, we have observed a
few cases of a limit cycle that is not circular as shown in
Fig.~\ref{fig:conf-trajs} a). Some chaotic confined trajectories can
also be found as shown in the bottom panel of
Fig.~\ref{fig:conf-trajs} b).

\section{Theoretical description of the vortex motion with rotating field}
\label{sec:field}

To describe analytically the observed vortex dynamics, a standard
procedure is to derive Thiele--like equations, as it was done in
Sec.~\ref{sec:nofield} without field. Due to the field there appears
the following Zeeman term in the total energy (see Appendix
\ref{sec:appendix-field} for details):
\begin{equation} \label{eq:V-eff}
V(\tau) \approx \pi bRL\cos\left(\Phi- \nu\tau\right).
\end{equation}
When the vortex reaches the limit cycle, the total energy is
constant. We have checked this fact in our simulations, calculating
the power--dissipation relation \eqref{eq:dH/dt}. For a vortex,
which moves according to the Thiele Ansatz, the power--dissipation
relation \eqref{eq:dE/dt} takes the form:
\begin{equation*}
\frac{d \mathscr{E}}{d\tau} = -\pi \eta\dot{\vec{R}}^2 + \pi b\nu
RL\sin\left(\Phi-\nu\tau\right).
\end{equation*}
The energy can tend to a constant value only when $\dot{\Phi}=\nu$,
so the frequency of the vortex motion should be equal to the driving
frequency. Thus the standard Thiele approach cannot provide the
circular motion of the vortex with the orbit frequency $\Omega<\nu$
we have observed in our simulations, see previous section. The
reason is that the field excites low--frequency quasi--Goldstone
modes \cite{Gaididei00,Zagorodny03}, which can couple with the
translation mode \cite{Kovalev03}. Therefore it is not correct to
describe the vortex as a rigid particle and it is necessary to take
into account the internal vortex structure.

To describe the approach to the limit cycle we now generalize the
collective variable theory to take into account an internal degree
of freedom of the vortex. Because the magnetic field changes the $z$
component of the magnetization and generates a new ground state, it
is natural to include into the $m$ field an additional degree of
freedom. To comply with the new ground state \eqref{eq:m0} we add to
the $\phi$--field \eqref{eq:Thiele-ansatz(2)} a time dependent phase
$\Psi(\tau)$ describing homogeneous spin precession. $\Psi(\tau)$
can be understood as the generalization of an arbitrary constant
phase which could be added in Eq.~\eqref{eq:Thiele-ansatz(2)}
without changing the dynamics. However, this constant phase does
influence the dynamics if there is a constant in--plane magnetic
field, which breaks the rotational symmetry in the xy--plane
\cite{Gouvea90}.

\begin{figure}
\includegraphics[width=\columnwidth]{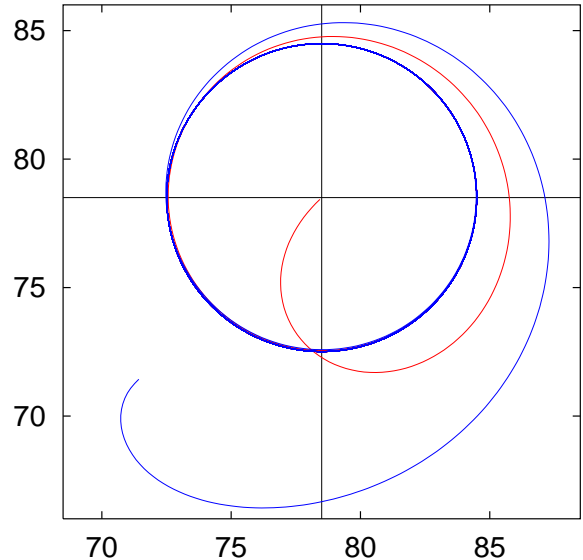}
\caption{ \label{fig:model} %
(Color online) Two trajectories of a vortex from the collective
variable Eqs.~\eqref{eq:EOM}, starting from different initial
positions. Red line: \mbox{$R(0)=a$}, \mbox{$\Phi(0)=5\pi/4$},
\mbox{$\Psi(0)=\pi/4$}. Blue line: \mbox{$R(0)=10a$},~\mbox{$\Phi(0)=
5\pi/4$}, \mbox{$\Psi(0)=\pi/4$}. Other parameters: $b=0.002$, $\nu =
0.125$, $\varepsilon=0.01$, and $\delta=0.08$.  System radius:
$L=78a\approx 44$.}
\end{figure}

The \emph{New Ansatz} that we choose is
\begin{subequations}  \label{eq:Ansatz}
\begin{align} \label{eq:Ansatz(m)}
m(z,\tau)   &= \cos \theta \left( \frac{\left| z-Z(\tau)\right|}{l(\tau)} \right),\\
\label{eq:Ansatz(phi)} %
\phi(z,\tau) &= \arg\Big(z-Z(\tau)\Big) - \arg\Big(z-Z_I(\tau)\Big)  \nonumber\\
            &\,+\arg Z(\tau)\,+\,\Psi(\tau),
\end{align}
\end{subequations}
which describes a mobile vortex structure like
\eqref{eq:Thiele-ansatz}, but including a precession of the spins as
a whole, through a time--dependent phase $\Psi(\tau)$ and a dynamics
of the vortex core, through the core width $l(\tau)$. The latter
allows a variation of the $z$--component of the magnetization. We
will see that in the Lagrangian the two variables $l$ and $\Psi$ are
conjugate to each other so that one needs to introduce them
together.

We find it convenient to use in the following, instead of $l(\tau)$,
the $z$--component of the total spin,
\begin{equation}   \label{eq:M}
M(\tau)  = \frac1\pi\int\!\! d^2 x\, m(z,\tau)  = M_0 l^2(\tau),
\end{equation}
which is related to the total number of ``spin deviations'' or
``magnons'', bound in the vortex. \cite{Kosevich90} Here $M_0$
\begin{equation} \label{eq:n0}
M_0 = 2  \int_0^{\infty}\!\! {\rho} d{\rho} \cos\theta({\rho})
\approx 2.75
\end{equation}
is related to the characteristic number of magnons bound in the
\emph{static} vortex. Note that without dissipation and for zero
field, $M$ is conserved. The field excites an internal dynamics,
changing the number of bound magnons and the total spin $M$.

To construct  effective equations we use the same variational
technique as in Section \ref{sec:nofield}. Besides the ``vortex
coordinates'' $\{R,\Phi\}$, we consider two ``internal variables''
$\{M,\Psi\}$ so that our set of collective variables is
\begin{equation} \label{eq:X_i}
X_i=\left\{R(\tau),\Phi(\tau),M(\tau),\Psi(\tau)\right\}.
\end{equation}
One can derive the effective Lagrangian of the system by inserting
ansatz \eqref{eq:Ansatz} into the full Lagrangian
\eqref{eq:Lagrangian}, and calculating the integrals, see Appendix
\ref{sec:appendix-field} for details:
\begin{widetext}
\begin{equation} \label{eq:L-eff-new}
\begin{split}
{\mathscr{L} \over \pi}= M \dot{\Psi}- R^2\dot{\Phi} -
\ln\frac{L^2-R^2}{L}
 + \frac12\left( \ln\frac{M}{M_0} - \frac{M}{M_0}\right)
- b LR f\left(\tfrac{R}{L}\right)\cos(\Phi+\Psi-\nu\tau).
\end{split}
\end{equation}
In the same way one can derive an effective dissipative function
\begin{equation}  \label{eq:F_diss-eff-new}
{\mathscr{F}\over \pi} = \frac{\varepsilon}{2}
\Biggl[\left(\dot{R}^2+R^2\dot{\Phi}^2\right)\left( C_1 + \frac12\ln
(L^2-R^2) - \frac12 \ln\frac{M}{M_0}\right) +
\dot{\Psi}^2\left(L^2-\frac{M}{M_0}\right) + 2
R^2\dot{\Phi}\dot{\Psi}  +  \frac{C_2 \dot{M}^2}{M M_0}\Biggr],
\end{equation}
where the constants $C_1$ and $C_2$ are introduced in \eqref{eq:C1}
and \eqref{eq:C2}, respectively. From the Euler--Lagrange equations
\eqref{eq:Euler-Lagrange} for the set of variables \eqref{eq:X_i} we
obtain finally
\begin{subequations}\label{eq:EOM}
\begin{align}
\label{eq:EOM-R} %
\dot{R}  &=  \frac{\varepsilon R}{2}  \left[ \dot{\Phi} \left(
C_1+\frac12\ln(L^2-R^2) -\frac12 \ln\frac{M}{M_0} \right) +
\dot{\Psi} \right] - \frac{b L}{2}f\left(\tfrac{R}{L}\right)
\sin\big( \Phi + \Psi - \nu \tau \big) ,
\\
\label{eq:EOM-Phi} %
\dot{\Phi}  &= \frac{1}{L^2-R^2} -\frac{\varepsilon \dot{R}}{2R}
\left( C_1+\frac12\ln (L^2-R^2) - \frac12 \ln\frac{M}{M_0} \right)
-\frac{b L}{2 R} g\left(\tfrac{R}{L}\right)\cos\big( \Phi+\Psi-\nu
\tau\big),
\\
\label{eq:EOM-M}%
\dot{M} &  = -\varepsilon \left[ R^2 \dot{\Phi} +
\dot{\Psi}\left(L^2-\frac{M}{M_0}\right) \right] + b L
Rf\left(\tfrac{R}{L}\right)\sin\big(\Phi+\Psi- \nu  \tau\big),
\\
\label{eq:EOM-Psi}%
\dot{\Psi} & = \frac12\left( \frac{1}{M_0}-\frac{1}{M}\right) +
\varepsilon C_2 \frac{\dot{M}}{MM_0}.
\end{align}
\end{subequations}
\end{widetext}

To integrate numerically the differential algebraic system
\eqref{eq:EOM}, one needs to solve at each step a linear system; we
used the MAPLE software \cite{maple} which includes such a facility.
The set of Eqs.~\eqref{eq:EOM} describes the main features of the
observed vortex dynamics, and yields the circular limit cycle for
the trajectory of the vortex center, see Fig.~\ref{fig:model}. Let
us note that Eqs.~\eqref{eq:EOM-R} and \eqref{eq:EOM-Phi} reduce to
the Thiele equations for the coordinates $(R,\Phi)$ of the vortex
center when $M$ and $\Psi$ are omitted and in this case no stable
closed orbit is possible. Only including the internal degrees of
freedom $(M,\Psi)$ one can obtain a stable limit cycle.

\begin{figure}
\includegraphics[width=\columnwidth]{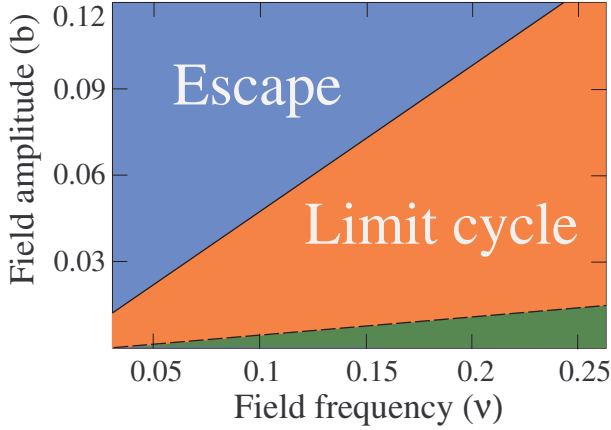}
\caption{ \label{fig:phase_diagram} %
(Color online) The two types of trajectories observed in the
$(\nu,b)$ field parameter plane for the collective variable
Eqs.~\eqref{eq:EOM}. The parameters are $\varepsilon=0.01$,
$\delta=0.08$ and system radius $L=36a\approx 20$.}
\end{figure}

In the parameter plane $(\nu,b)$ shown in
Fig.~\ref{fig:phase_diagram} we indicate the two main types of
trajectories found by numerical integrating  Eqs.~\eqref{eq:EOM}.
Vortex trajectories converge to a limit cycle only for $b\lesssim
\nu/2$ (red domain). When the amplitude of the rotating field $b$
lies above the critical curve, the vortex escapes from the system
along a spiral trajectory (blue domain). The model has no lower
boundary for the limit cycle. However when the amplitude of the
field lies below the critical curve (dashed line in
Fig.~\ref{fig:phase_diagram}), the radius of the vortex orbit can
become less than the lattice constant (green domain). In this case
discreteness effects are important for the spin system, so the model
can no longer be adequate.

\begin{figure}
\includegraphics[width=\columnwidth]{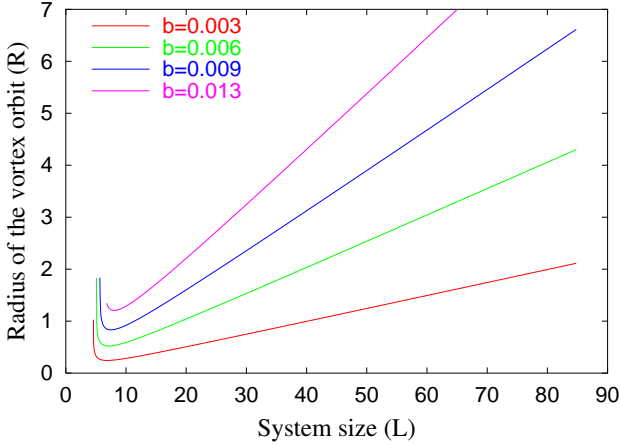}
\caption{ \label{fig:R_L_model} %
(Color online) Radius of the vortex orbit $R$ \emph{vs.} the system
size $L$, in the circular limit cycle for the collective variable
Eqs.~\eqref{eq:EOM} for a field frequency $\nu=0.06$. The other
parameters are the same as in Fig.~\ref{fig:phase_diagram}.}
\end{figure}

In Fig.~\ref{fig:R_L_model} we show the radius of the vortex orbit
$R$ on the circular limit cycle as a function of the system size
$L$, obtained from the numerical solution of Eqs.~\eqref{eq:EOM}.
Notice the linear dependence $R\propto L$ similar to the one
observed in the numerical simulations (see Fig.~\ref{fig:Rv-L}).

To analyze the main features of the model we simplify it, assuming
that the vortex orbit is never close to the system border ($R\ll L$)
and that the total $z$--component of the spin varies weakly so that
$N\equiv (M-M_0)/M_0\ll 1$. Then one can simplify the expressions
for the Lagrangian and dissipative function where the common factor
$\pi$ has been omitted:
\begin{subequations} \label{eq:L&F-simple}
\begin{align}
\label{eq:L-simple} %
\mathscr{L} &= M_0 N\dot{\Psi} - R^2\dot{\Phi} + {R^2\over L^2}
- \frac{N^2}{4} - bRL\cos \Delta ,\\
\label{eq:F_diss-simple} %
\mathscr{F} &= \eta\left(\dot{R}^2+R^2\dot{\Phi}^2\right) +
\varepsilon \frac{L^2}{2}\dot{\Psi}^2 + \varepsilon R^2
\dot{\Phi}\dot{\Psi},
\end{align}
\end{subequations}
where $\Delta \equiv \Phi + \Psi - \nu\tau$, and $\eta$ was defined
in \eqref{eq:eta}.

The equations of motion which result from \eqref{eq:L&F-simple} have
the simple form:
\begin{subequations} \label{eq:simple}
\begin{align}
\label{eq:simple-P} %
\dot{R} &= \eta R \dot{\Phi}
- \frac{bL}{2}\sin \Delta  +\varepsilon {R\over 2} \dot{\Psi},\\
\label{eq:simple-Phi} %
\dot{\Phi} & = {1\over L^2}  - \eta {\dot{R} \over R}
- \frac{bL}{2R }\cos \Delta ,\\
\label{eq:simple-N} %
M_0\dot{N} &= - \varepsilon L^2 \dot{\Psi} + bRL\sin\Delta
- \varepsilon R^2 \dot{\Phi},\\
\label{eq:simple-Psi} %
2M_0\dot{\Psi} &= N,
\end{align}
\end{subequations}

The set of Eqs.~\eqref{eq:simple} describes two damped periodically
forced oscillators, described by two couples of variables,
$(R,\Phi)$ and $(N,\Psi)$. Under the action of forcing these
oscillators can phase-lock and induce the limit cycle. The numerical
study of Eqs.~\eqref{eq:simple} reveals three different types of
behaviors as a function of the field amplitude $b$ for a fixed
frequency $\nu$. We choose $\nu=0.06$. For a small $b=3\cdot
10^{-4}$, the phase $\Delta$ increases linearly with time, $N$
oscillates and $R$ increases very slowly without stabilization. When
the amplitude is large such as $b=0.12$, $\Delta$ tends to $-\pi$,
$N$ becomes negative and then goes back to about $0$, $R$ increases
indefinitely. For $b=3\cdot 10^{-4}$, $N$ tends to a positive
constant, $\Delta$ tends to $\pi$ so that terms in $\dot R$ balance
and we have the limit cycle. One can see that the dynamics of the
couple $(N, \Psi)$ is fast with a typical relaxation time of about
$1/\varepsilon L^2$ while the dynamics of the couple $(R,\Phi)$ is
slow and depends on the initial position $R_0$. The limit cycle is
obtained for $R_0<0.6 L$, outside that range $R$ increases
indefinitely.

When the solution of the system of Eqs.~\eqref{eq:simple} converges
to a limit cycle, we have
\begin{equation} \label{eq:Phi-Psi-relation}
\dot{R} = \dot{N} = 0,\qquad  \dot{\Phi}\equiv \Omega =
\text{const}, \quad \dot{\Psi} = \nu -\Omega.
\end{equation}
In that case we obtain the following three algebraic equations
\begin{subequations} \label{eq:limitcycle}
\begin{align}
\label{eq:lc1} %
 2R(\nu+A\Omega) &= bL\sin\Delta,\\
\label{eq:lc2} %
\varepsilon L(\nu-\Omega) &= bR\sin\Delta,\\
\label{eq:lc3} %
-2R\Omega &= bL\cos\Delta,
\end{align}
\end{subequations}
where $A=C_1-1+\ln L$. Extracting the $\sin\Delta$ term from the
first and second equation, we obtain the frequency of the vortex
motion
\begin{equation} \label{eq:Omega1}
\Omega \approx \frac{\nu}{1+AR^2/L^2}.
\end{equation}
We now eliminate the sine and cosine terms from \eqref{eq:lc1} and
\eqref{eq:lc3}, resulting in $R\approx bL/2\Omega$. Combining with
\eqref{eq:Omega1} one has
\begin{equation} \label{eq:Omega}
\Omega \approx \frac{\nu+\sqrt{\nu^2 - Ab^2}}{2}.
\end{equation}
This value is smaller than the driving frequency $\nu$ in accordance
with our simulations. However, it is not proportional to $1/L$ as in
the spin simulations.
For the radius of the limit cycle we have finally
\begin{equation} \label{eq:R}
R\approx \frac{bL}{\nu + \sqrt{\nu^2-Ab^2}}\approx \frac{bL}{2\nu}.
\end{equation}
The radius of the vortex orbit $R$ depends linearly on the system
size in good agreement with the results of the simulation, see
Section \ref{sec:simulations}. It also bears the proportionality to
$1/\nu$ observed in the spin dynamics.

The range of parameters, which admits limit cycle trajectories, can
be estimated from the natural condition $R < L$, which gives $b <
2\nu$. However, there exist stronger restrictions for the limit
cycle. The solution \eqref{eq:Omega} is real (not complex) only when
$\nu^2 -Ab^2>0$. Another limit for the parameters is obtained from
the natural condition $Rl_0>a$ (discreteness effects are important
there). Thus the range of parameters, which admits the limit cycle
trajectories can be estimated as follows:
\begin{equation} \label{eq:phase-diagram}
\begin{split}
\frac{2a}{l_0L} < \frac{b}{\nu} < \varkappa
\end{split}
\end{equation}
with $\varkappa = 1/\sqrt{A} = 1/\sqrt{C_1-1+\ln L}$.

For the parameters considered in Fig.~\ref{fig:phase_diagram}
$\varkappa\approx 0.48$ so that the estimate
\eqref{eq:phase-diagram} agrees with the boundary $b\approx\nu/2$
shown in the figure.

From the above expressions one can estimate $\dot{\Psi}$ on the
limit cycle as
\begin{equation*}
\dot{\Psi} \approx \frac{\nu-\sqrt{\nu^2 - Ab^2}}{2},
\end{equation*}
which shows that the change in magnetization $N = 2M_0\dot{\Psi}$
due to the internal variables is small. It is nevertheless crucial
for obtaining the limit cycle.

\section{Discussion}
\label{sec:discussion}

\begin{figure}
\centering
\includegraphics[width=\linewidth]{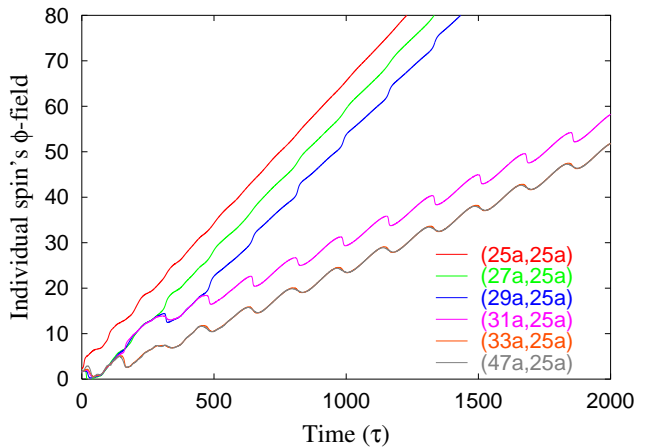}
\caption{(Color online) Time evolution of the $\phi$--field for
spins inside and outside the vortex orbit, once the vortex has
reached a  circular limit cycle. The parameters are $\nu=0.1$,
$b=0.02$, and $L=48a\approx 27$. Inside spins ares located at
$(25a,25a)$, $(27a,25a)$, $(29a,25a)$, and the outside spins are at
$(31a,25a)$, $(33a,25a)$, $(47a,25a)$. \label{fig:phi}}
\end{figure}

Another way to understand the vortex dynamics is to analyze the
movement of individual spins. In a set of simulations, we recorded
the components of some individual spins to observe their time
evolution. We consider a large enough time so that the vortex
reaches the limit cycle. For the Fourier spectrum of the
$z$--component of individual spins we have observed some peaks,
which appear naturally with the frequency of the limit cycle
$\Omega$. Every time the vortex passes close to the spins, the spins
feel a lick upwards. The behavior of $\phi(\tau)$ for several spins
is shown in Fig.~\ref{fig:phi}. When the vortex has reached its
limit cycle ie for $t>500$ the spins behave differently whether they
are inside or outside the vortex orbit. Inside,  $\phi$ is quite
regular and increases linearly with time at a rate given by $\nu$,
with $\phi \approx \varphi_0+\nu\tau$. This is shown by the three
upper curves in Fig.~\ref{fig:phi} for $\tau> 500$ which is the time
taken by the vortex to settle on its orbit. Outside the orbit and
for $\tau> 500$, the increase of $\phi$ is more irregular as shown
by the three lower curves in Fig.~\ref{fig:phi}. There the Fourier
spectrum of $\phi(\tau)$ has a main frequency $\omega - \Omega$
together with additional peaks at $\omega \pm n\Omega$ where $n$ is
an integer.

Our collective variable theory describes this effect as we show now.
We assume that the vortex has reached the limit cycle so that the
variables $\Phi$ and $\Psi$ fulfil the relations
\eqref{eq:Phi-Psi-relation}. According to the Ansatz
\eqref{eq:Ansatz(phi)}, on the limit cycle the dynamical variable
$\phi$ can be written as
\begin{equation} \label{eq:Ansatz(phi)1} %
\phi(z,\tau) =  \varphi_0 + \nu\tau + \arg\Big(z-Z(\tau)\Big) -
\arg\Big(z-Z_I(\tau)\Big).
\end{equation}
We consider the vortex to be far from the boundary, i.e. $R\ll L$.
Then the radius of the image--vortex trajectory is $R_I=L^2/R\gg L$,
and for any $|z|<L$ the last term in Eq.~\eqref{eq:Ansatz(phi)1}
$\arg(z-Z_I(\tau))\approx \pi + \Phi(\tau)$, so
\begin{equation} \label{eq:Ansatz(phi)2} %
\phi(z,\tau) = \tilde{\varphi}_0 +(\nu -\Omega)\tau +
\arg\Big(z-Z(\tau)\Big).
\end{equation}
If we consider a spin, situated at a distance $|z|>R$, the last term
in Eq.~\eqref{eq:Ansatz(phi)2} describes only small oscillations on
the background of the main dependence $\phi(z,\tau) =
\tilde{\varphi}_0 +(\nu -\Omega)\tau$. At the same time for a spin
located at $|z|<R$, this term is decisive. Let us consider the
limiting case of a spin situated near the center of the system. Then
$\arg(z-Z(\tau))\approx \pi + \Phi(\tau)$, and
Eq.~\eqref{eq:Ansatz(phi)2} can be simply written as $\phi(z,\tau) =
\varphi_0 + \nu\tau$. Thus, the two regimes for the in--plane
components of the spins  are well--pronounced, which is confirmed by
our simulations, see Fig.~\ref{fig:phi}.

In a wide range of parameters the vortex moves along a limit
circular trajectory. When the intensity of the ac field exceeds a
critical value, $b>\varkappa\nu$, the vortex escapes through the
boundary and annihilates. This process is important for practical
applications, because vortices are known to cause hysteresis loop in
magnetic nanostructures \cite{Cowburn02}. Usually static fields are
considered in the experiments and these cause a hysteresis of the
$M_x(H_x)$ loop, see e.g.
Refs.~\onlinecite{Cowburn99,Fernandez00,Lebib01,Schneider02}. The
saturation field in the static regime to obtain a hysteresis is
about $\omega_0/\gamma$ (in dimensionless units $b\sim1$). In this
article we consider an ac driving of the vortex, which causes a
dynamical hysteresis, $M_x$ as a function of the intensity of the ac
field $b$. Typical fields for vortex annihilation, $b\sim
\varkappa\nu\ll1$, are much weaker than in the static regime. It is
then much easier to destabilize the vortex with an ac field than
with a dc field.

Let us make some estimates. We choose permalloy (Py,
$Ni_{80}Fe_{20}$) magnetic nanodots \cite{Cowburn99,Schneider02}.
The measured value of $M_s=\gamma S L^2/a^2 = 770\,G$, the exchange
constant $A = JS^2=1.3\times 10^{-6}\,erg/cm$, and $\gamma/2\pi =
2.95\,GHz/kOe$. \cite{Park03} Typical fields of the vortex
annihilation $b\sim \varkappa\nu$, which is about some tens of $Oe$.

Another important fact can be seen from Fig.~\ref{fig:R_L_model}:
the vortex is unstable in small magnetic dots, the typical minimal
size $L_{\text{min}}\sim 5$. For the Py magnetic dot with the
magnetic length $l_0=5.9\,nm$, \cite{Park03} the minimal size for
the vortex state magnetic dot under weak ac driving is about
$L_{\text{min}} l_0\sim30\,nm$. This means that for magnetic dots
with diameters less than $60\,nm$ the vortex state is unstable
against the ac field giving rise to a single--domain state.

In conclusion, we developed a new collective variables approach
which describes the vortex dynamics under a periodic driving, taking
into account internal degrees of freedom. To our knowledge, it is
the first time that an interplay between internal and external
degrees of freedom, giving raise to the existence of stable
trajectories, is observed in the case of 2D magnetic structures.
This ansatz gives (up to a factor of 2) the radius of the limit
cycle. Also the dependencies of $R$ on the system size $L$, the
field amplitude, and the frequency are correct. However, the
dependence of the vortex orbit frequency $\Omega$ on the system size
is different from the one in the spin dynamics. Moreover, in the
collective variable theory the magnetization and vortex position
variables vary on very different time scales, this is not the case
for the spin dynamics. Despite this we think that this collective
variable approach is very general and can be employed for the
self--consistent description of the dynamics of different 2D
nonlinear excitations, e.g. topological solitons in 2D easy--axis
magnets \cite{Sheka01}.

%
%

\acknowledgments

F.G.M. and J.G.C. acknowledge support from a French--German Procope
grant (nb 04555TG). Part of the computations was done at the Centre
de Ressources Informatiques de Haute--Normandie. D.D.Sh. and Yu.G.
thank the University of Bayreuth, where part of this work was
performed, for kind hospitality and acknowledge support from
Deutsches Zentrum f{\"u}r Luft- und Raumfart e.V., Internationales
B{\"u}ro des Bundesministeriums f{\"u}r Forschung und Technologie,
Bonn, in the frame of a bilateral scientific cooperation between
Ukraine and Germany, project No. UKR--02--011. J.P.Z. is supported
by a grant from Deutsche Forschungsgemeinschaft.

%
%
%
%

\appendix

\section{Discrete spin dynamics}
\label{sec:appendix-Discrete}

While Eqs. \eqref{eq:LL-discrete} are convenient for analytical
consideration the presence of the time derivative on both sides
makes them inconvenient for numerical simulations. Equivalent
equations are obtained by forming the cross product of
\eqref{eq:LL-discrete} with $\vec{S}_{\vec{n}}$ and subtracting the
result from \eqref{eq:LL-discrete}. In this way we get
\begin{equation} \label{eq:LL-discrete(num)}
\begin{split}
(1 + \varepsilon^2) \frac{d \vec{S}_{\vec{n}} }{dt}=
\left[\vec{S}_{\vec{n}} \times \vec{F}_{\vec{n}}\right] -
\frac{\varepsilon}{S} \left[ \vec{S}_{\vec{n}}\times
\left[\vec{S}_{\vec{n}} \times \vec{F}_{\vec{n}}\right]\right],
\end{split}
\end{equation}
where $\vec{F}_{\vec{n}}=-{\partial \mathscr{H} }/{\partial
\vec{S}_{\vec{n}}}$ is the total effective field; the factor $(1 +
\varepsilon^2)$ is usually neglected, or absorbed into
$\mathscr{H}$, giving effective constants $J$, $K$ and $B$.

From the discrete dynamics \eqref{eq:LL-discrete(num)} one easily
derives the power--dissipation relation for the total energy
$\mathscr{H}=-\sum_{\vec{n}}\vec{S}_{\vec{n}}
\cdot\vec{F}_{\vec{n}}$. We have
\begin{align*}
\frac{d \mathscr{H}}{d t} &= - \sum_{\vec{n}} \vec{S}_{\vec{n}}
\cdot \frac{d \vec{B}}{d t} - \sum_{\vec{n}} \vec{F}_{\vec{n}} \cdot
\frac{d \vec{S}_{\vec{n}}}{d t}
=- \sum_{\vec{n}} \vec{S}_{\vec{n}} \cdot \frac{d \vec{B}}{d t}\\
&+ \frac{\varepsilon}{(1 + \varepsilon^2)S} \sum_{\vec{n}}
\vec{F}_{\vec{n}} \cdot \left[ \vec{S}_{\vec{n}} \times  \left[
\vec{S}_{\vec{n}} \times \vec{F}_{\vec{n}}\right]\right]
\end{align*}
and finally
\begin{equation} \label{eq:dH/dt}
 \frac{d \mathscr{H}}{d t} =  - \frac{\varepsilon}{(1 + \varepsilon^2)S}
\sum_{\vec{n}} \left[ \vec{S}_{\vec{n}} \times \vec{F}_{\vec{n}}
\right]^2 - \sum_{\vec{n}} \vec{S}_{\vec{n}} \cdot \frac{d
\vec{B}}{d t}.
\end{equation}
While the first term is always negative, it is the second term which
can give rise to transients in the relaxation to equilibrium, or
even the resonances, depending on the parameters of the
time-dependent magnetic field.

\section{Collective Variable Equations without field}
\label{sec:appendix-nofield}

\begin{figure}
\includegraphics[width=\linewidth]{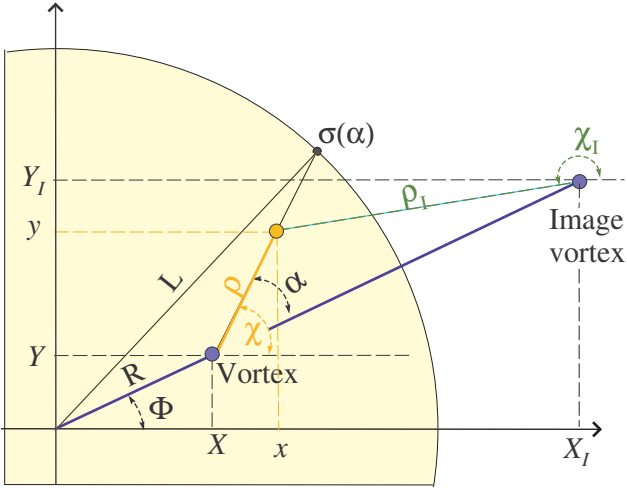}
\caption{ (Color online) Arrangement of angles in the mobile frame
centered in the vortex with the coordinates $Z=X+iY=R\exp(i\Phi)$.}
\label{fig:mobile-frame}
\end{figure}

It is convenient to make calculations in the reference frame
centered on the vortex whose axes are parallel to the standard frame
\begin{equation} \label{eq:moving-frame}
x - X(\tau) = \rho \cos\chi,\quad  y - Y(\tau) = \rho \sin\chi ~~.
\end{equation}
Viewed from this point, the distance to the circular border of the
system changes as a function of the azimuthal angle $\chi$, see
Fig.~\ref{fig:mobile-frame}. Every integral over the domain $|z|<L$
can then be calculated as
\begin{equation*}
\int\limits_{|z|<L}\! f(\rho,\alpha) d^2\xi = \bigl\langle
F(\alpha)\bigr\rangle, \quad
F(\alpha)=2\pi\!\!\int\limits_0^{\sigma(\alpha)}\!
f(\rho,\alpha)\rho d\rho,
\end{equation*}
where $\alpha = \chi-\Phi$ is given by the cosine theorem
\begin{equation}
\sigma(\alpha) = -R \cos\alpha + \sqrt{L^2 - R^2 \sin^2\alpha},
\end{equation}
and the averaging means $\bigl\langle
F(\alpha)\bigr\rangle=\frac{1}{2\pi}\int_0^{2\pi}F(\alpha)d\alpha$.
We also have the relations
\begin{subequations}  \label{eq:orts}
\begin{align}
\label{eq:orts(1)} %
\vec{e}_{\rho} &=~~\,\vec{e}_{x}\,\cos\chi + \vec{e}_{y}\,\sin\chi,
\\
\label{eq:orts(2)} %
\vec{e}_{\chi} &= -\,\vec{e}_{x}\,\sin\chi + \vec{e}_{y}\,\cos\chi,
\\
\label{eq:orts(3)} %
\vec{e}_{R} &=~~\,\vec{e}_{x}\,\cos\Phi + \vec{e}_{y}\,\sin\Phi,
\\
\label{eq:orts(4)} %
\vec{e}_{\Phi} &= -\,\vec{e}_{x}\,\sin\Phi + \vec{e}_{y}\,\cos\Phi.
\end{align}
\end{subequations}

In order to derive an effective Lagrangian we start with the
``microscopic'' Lagrangian \eqref{eq:Lagrangian},
\begin{equation} \label{eq:L-via-G}
\mathscr{L} = \mathscr{G} - \mathscr{E}^{\text{int}}, \qquad
\mathscr{G} = -\int d^2\xi  (1-m)\dot{\phi}.
\end{equation}
We will provide all the calculations for the vortex with unit
vorticity $q=1$ and positive polarity $p=1$. Using the traveling
wave Ansatz \eqref{eq:Thiele-ansatz(2)} in the form,
\begin{equation*}
\phi(z,\tau) = \chi - \chi_I + \Phi,
\end{equation*}
one can calculate the time derivatives in the moving frame
\eqref{eq:moving-frame}:
\begin{equation*}
\begin{split}
\dot{\chi}    &= \frac{\dot{R}}{\rho} \sin\alpha - \frac{R\dot{\Phi}}{\rho}\cos\alpha.\\
\dot{\chi}_I &=  - \frac{L^2\rho\dot{R}}{R^2\rho_I^2} \sin\alpha +
\frac{L^2\dot{\Phi}}{R\rho_I^2}\sqrt{\rho_I^2-\rho^2\sin^2\alpha}.
\end{split}
\end{equation*}
Here $\rho_I = \left\lvert z-Z_I(\tau) \right\rvert$,
$\chi_I=\arg\left(z - Z_I(\tau)\right)$. In the main approach for
$R/L\ll1$, one can simplify an expression for $\dot{\chi}_I$, so
finally we have
\begin{equation} \label{eq:dotphi-nofield}
\begin{split}
\dot{\phi}    &= \left(\dot{R} \sin\alpha -
R\dot{\Phi}\cos\alpha\right) \left(\frac{1}{\rho} +
\frac{\rho}{L^2}\right).
\end{split}
\end{equation}

Then the gyroterm in the Lagrangian $\mathscr{G}$ gives
\begin{equation*} \label{eq:G-nofield}
\begin{split}
\mathscr{G}    = & \mathscr{G}_1+\mathscr{G}_2,\\
\mathscr{G}_1 = & -\int d^2\xi  \dot{\phi} =
- 2\pi\dot{R}\Bigl\langle[\sigma(\alpha)-L/3]\sin\alpha\Bigr\rangle\\
&+ 2\pi R\dot{\Phi}\Bigl\langle[\sigma(\alpha)-L/3]\cos\alpha\Bigr\rangle ,\\
\mathscr{G}_2 = & \int d^2\xi m \dot{\phi} =
  k_0\dot{R}\bigl\langle\sin\alpha\bigr\rangle
- k_0R\dot{\Phi}\bigl\langle\cos\alpha\bigr\rangle,
\end{split}
\end{equation*}
where the constant $k_0 = 2\pi \int_0^{\infty}\cos\theta({\rho})
d{\rho}$. After averaging with account of the expressions
\begin{equation} \label{eq:<sigma>}
\Bigl\langle\sigma(\alpha)\sin\alpha\Bigr\rangle = 0,\quad
\Bigl\langle\sigma(\alpha)\cos\alpha\Bigr\rangle = -\frac{R}{2},
\end{equation}
we obtain the gyroterm in the form
\begin{equation} \label{eq:G-final-nofield}
\mathscr{G}_1 = - \pi R^2\dot{\Phi}, \qquad \mathscr{G}_2=0,
\end{equation}
and finally, $\mathscr{G} = -\pi R^2\dot{\Phi}$.

Let us calculate an effective dissipative function, starting from
the ``microscopic'' dissipative function \eqref{eq:F-diss}, which we
cut into two terms, $\mathscr{F}=\mathscr{F}_1+\mathscr{F}_2$ with
\begin{equation*} \label{eq:Fidiss_1-and-2}
\mathscr{F}_1 = \frac{\varepsilon}{2} \int\! d^2\xi\
\frac{{\dot{m}}^2}{1-m^2}, \quad \mathscr{F}_2 = \frac{\varepsilon
}{2} \int\! d^2\xi\ (1-m^2){\dot{\phi}}^2.
\end{equation*}
The time derivative of the $m$--field  can be easily calculated in
the moving frame \eqref{eq:moving-frame}, using the traveling wave
Ansatz \eqref{eq:Thiele-ansatz}
\begin{equation} \label{eq:dotm-nofield}
\dot{m} = \theta^\prime\sin\theta\left(\dot{R} \cos\alpha +
R\dot{\Phi}\sin\alpha\right).
\end{equation}
Calculating integrals for $\mathscr{F}_1$ with account of
\eqref{eq:dotm-nofield}, we derive:
\begin{equation} \label{eq:F1}
\mathscr{F}_1 = \frac{\varepsilon
\pi}{2}k_1\left(\dot{R}^2+R^2\dot{\Phi}^2\right),
\end{equation}
where $k_1=\int_0^\infty{\theta}^{\prime2}({\rho}) {\rho} d{\rho}$.
In the same way we can derive $\mathscr{F}_2$, taking into account
$\dot{\phi}$ from Eq.~\eqref{eq:dotphi-nofield},
\begin{equation} \label{eq:F2}
\begin{split}
\mathscr{F}_2 &= {\varepsilon \pi} \Biggl(
\dot{R}^2\Bigl\langle(k_2+\ln\sigma)\sin^2\alpha\Bigr\rangle\\
&+R^2\dot{\Phi}^2\Bigl\langle(k_2+\ln\sigma)\cos^2\alpha\Bigr\rangle\\
&-R\dot{R}\dot{\Phi}\Bigl\langle(k_2+\ln\sigma)\sin2\alpha\Bigr\rangle
\Biggr),\\
k_2&=\frac54+\int_0^1\frac{\sin^2\theta({\rho})}{{\rho}} d{\rho} -
\int_1^\infty\frac{\cos^2\theta({\rho})}{{\rho}} d{\rho}.
\end{split}
\end{equation}
Using the averages
\begin{equation} \label{eq:<ln(sigma)>}
\begin{split}
&\Bigl\langle\sin^2\!\alpha\ \ln\sigma(\alpha)\Bigr\rangle
    = \Bigl\langle\cos^2\!\alpha\ \ln\sigma(\alpha)\Bigr\rangle\\
&  = \frac12\Bigl\langle\ln\sigma(\alpha)\Bigr\rangle
    = \frac14\ln\left(L^2-R^2\right)
\end{split}
\end{equation}
we calculate the dissipative function in the form
\begin{equation}  \label{eq:F_diss-nofield}
\mathscr{F} = \frac{\varepsilon\pi}{2}\left[C_1 + \frac12 \ln
(L^2-R^2)\right] \left(\dot{R}^2+R^2\dot{\Phi}^2\right).
\end{equation}
Here the constant $C_1=k_1+k_2$,
\begin{equation} \label{eq:C1}
\begin{split}
C_1 &= \frac54 + \int_0^1\frac{\sin^2\theta({\rho})}{{\rho}} d{\rho}
- \int_1^\infty\frac{\cos^2\theta({\rho})}{{\rho}} d{\rho}\\
& + \int_0^\infty {\theta}^{\prime2}({\rho}) {\rho} d{\rho} \approx
2.31.
\end{split}
\end{equation}
Supposing that the vortex is not close to the boundary, i.e. $R\ll
L$, we obtain the effective dissipative function in the form
\eqref{eq:F-eff}.

\section{Collective Variable Equations with field}
\label{sec:appendix-field}

First we calculate an effective Zeeman energy for the standard
Thiele--like motion of the vortex. Inserting the traveling wave
Ansatz \eqref{eq:Thiele-ansatz} into the ``microscopic'' Zeeman
energy \eqref{eq:V-continuum}, and calculating the integrals, we get
the effective energy in the form:
\begin{align} \label{eq:V-old-ansatz}
V(\tau) &= -\frac12b\int_0^{2\pi}d\chi \left[\sigma^2(\chi-\Phi) - c_1 \right]\cos(\phi - \nu\tau)\nonumber\\
&=\pi bRL f\left(\tfrac{R}{L}\right)\cos\left(\Phi - \nu\tau\right),
\end{align}
where
\begin{equation} \label{eq:f(x)}
f(x) = \frac{4}{3\pi}\left[E(x)\left(\frac{1}{x^2}+1\right)  -
K(x)\left(\frac{1}{x^2}-1\right)\right],
\end{equation}
where $E(x)$ and $K(x)$ are elliptical integrals. When the vortex is
far from the boundary, which is the case of interest, one can expand
this function into the series, $f(x) \approx 1-{x^2}/{8}$. In the
main approach it leads to the Zeeman term in the form
\eqref{eq:V-eff}. The corresponding magnetic force
\begin{equation} \label{eq:F_h-exact}
\begin{split}
\vec{F}_h = -\vec{\nabla}_{\!\!\vec{R}} V &=  \vec{e}_{\! R}\, \pi bLg\left(\tfrac{R}{L}\right)\cos\left(\Phi - \nu\tau\right)\\
&- \vec{e}_{\!\chi}\, \pi
bRLf\left(\tfrac{R}{L}\right)\sin\left(\Phi - \nu\tau\right),
\end{split}
\end{equation}
where the function
\begin{equation} \label{eq:g(x)}
\begin{split}
g(x) &= f(x)+xf'(x) \\
&= \frac{4}{3\pi}\left[K(x)\left(\frac{1}{x^2}-1\right)  -
E(x)\left(\frac{1}{x^2}-2\right)\right].
\end{split}
\end{equation}
For $x\ll1$ it has the following expansion $g(x) \approx
1-{3x^2}/{8}$.

Let us calculate the same Zeeman energy using the new Ansatz
\eqref{eq:Ansatz}. One can derive a Zeeman term similar to
\eqref{eq:V-old-ansatz}
\begin{equation} \label{eq:V-new-ansatz}
V(\tau) = \pi  bRL
f\left(\tfrac{R}{L}\right)\cos\left(\Phi+\Psi-\nu\tau\right).
\end{equation}
Besides this direct influence on the system, the magnetic field also
changes the gyroterm in the effective Lagrangian, and the energy of
the system. These changes result from the internal motion of the
vortex through $l(\tau)$, and from the uniform spin precession
through $\Psi(\tau)$. This does not change the gyroterm
$\mathscr{G}_1$, which has the same form as in
\eqref{eq:G-final-nofield}, but there appears the contribution
$\mathscr{G}_2=M\dot{\Psi}$. This can be easily calculated with
account of the time derivative
\begin{equation} \label{eq:dotphi-field}
\dot{\phi} = \dot{\Psi} + \left(\dot{R} \sin\alpha -
R\dot{\Phi}\cos\alpha\right) \left(\frac{1}{\rho} +
\frac{\rho}{L^2}\right).
\end{equation}
The total energy functional \eqref{eq:E-total} can be written in the
form $\mathscr{E}=\mathscr{E}_1+\mathscr{E}_2+\mathscr{E}_3+V$ with
\begin{subequations}  \label{eq:E}
\begin{align}
\label{eq:E(1)} \mathscr{E}_1 &= \frac12\int d^2\xi
\frac{(\vec{\nabla}m)^2}{1-m^2} = k_1 \pi,
\\
\label{eq:E(2)} \mathscr{E}_2 &= \frac12\int d^2\xi
(1-m^2)(\vec{\nabla}\phi)^2 \approx \pi \ln\frac{L^2-R^2}{l(\tau)L},
\\
\label{eq:E(3)} \mathscr{E}_3 &= \frac12\int d^2\xi\ m^2 = \frac{\pi
l^2(\tau)}{2}.
\end{align}
\end{subequations}
The term $\mathscr{E}_2$, which describes the interaction between
the vortex and its image, can be derived from \eqref{eq:E-int},
simply replacing $l_0$ by $l(\tau)$. In the last anisotropy term
$\mathscr{E}_3$ we have used the relation $\int_0^\infty
\cos{\theta}^2({\rho}) {\rho} d{\rho} = 1/2$, see
Ref.~\onlinecite{Ivanov95b}. Combining all terms of the Lagrangian
and omitting the constant term $\mathscr{E}_1$, one obtains the
effective Lagrangian of the system \eqref{eq:L-eff-new}.

The dissipative function contains two dynamical contributions. The
first one is due to the time dependence of the $m$--field:
\begin{equation} \label{eq:dotm-field}
\dot{m} = \frac{\theta^\prime\sin\theta}{l(\tau)}
\left(\frac{\dot{M}}{2M} \rho +\dot{R} \cos\alpha +
R\dot{\Phi}\sin\alpha \right).
\end{equation}
This term $\mathscr{F}_1$ can be derived in way similar to
\eqref{eq:F1}:
\begin{align}
\label{eq:F1-field} \mathscr{F}_1& = \frac{\varepsilon
\pi}{2}\left(k_1\dot{R}^2+k_1R^2\dot{\Phi}^2 + \frac{C_2
{\dot{M}}^2}{M M_0}\right),
\\
\label{eq:C2} C_2 &= \frac12  \int_0^\infty
{\theta}^{\prime2}({\rho}) {\rho}^3 d{\rho}  \approx 0.48.
\end{align}
To calculate the second term $\mathscr{F}_2$ we use $\dot{\phi}$
from Eq.~\eqref{eq:dotphi-field} and obtain
\begin{equation} \label{eq:F2-field}
\begin{split}
\mathscr{F}_2 = \frac{\varepsilon\pi}{2}  \Biggl\{ &
\left(\dot{R}^2+R^2\dot{\Phi}^2\right)
\left[k_2 + \frac12\ln \frac{L^2-R^2}{l^2(\tau)}\right]\\
&+ \dot{\Psi}^2\left[L^2-l^2(\tau)\right] + 2
R^2\dot{\Phi}\dot{\Psi}\Biggr\}.
\end{split}
\end{equation}
The total effective dissipative function $\mathscr{F} =
\mathscr{F}_1 + \mathscr{F}_2$ has the form
\eqref{eq:F_diss-eff-new}.


\end{document}